\documentclass[a4,12pt]{article}
\begin{document}

\setlength{\parindent}{15pt}
\setlength{\baselineskip}{18pt}

\def\thefootnote{\fnsymbol{footnote}}

\noindent  
{\bf \large A Quantum Theoretical Formulation of Proper Time for Spin $1/2$ Particles}

\vskip 5mm

\noindent
Shoju Kudaka $^1$ and Shuichi Matsumoto $^2$\footnote{Corresponding author.\\ 
{\it E-mail address}: shuichi@edu.u-ryukyu.ac.jp (S. Matsumoto).}\\
{\it $^1$Department of Physics, University of the Ryukyus, Okinawa 903-0213, 
Japan}
{\it $^2$Department of Mathematics, University of the Ryukyus, Okinawa 903-0213, Japan}

\vskip 1cm
\noindent
{\bf Abstract} 

We find a quantum mechanical formulation of proper time for spin $1/2$ particles within the framework of the Dirac theory. It is shown that an operator corresponds to the rate of the proper time and that the operator contains terms which oscillate with a very high frequency. We deduce, as an effect of these terms, the existence of an interference between the magnetic field and the rate of proper time. There is a possibility that the conclusion derived in this letter has some implications for astrophysics. 

\vskip 1cm

Many physicists have tried to elucidate the role that time should play in quantum mechanics. The fundamental question is whether time is an operator or merely a parameter. This question has occupied physicists since quantum mechanics was established early in the last century\ \cite{Jam}, and various approaches have been made to the problem. There has been much argument about the coordinate time and Heisenberg's time-energy uncertainty relation\ \cite{Jam, AhaMa}; some authors try to associate an operator with the coordinate time and then discuss related issues\ \cite{Grot, Wang}. On the other hand, the proper time of a particle has also received much attention; it is generally claimed that there exists an uncertainty relation between the proper time and the rest mass\ \cite{Gr,Un,Aha-2,Burakov}. In recent papers we have argued that the very act of weighing a particle interferes with the rate of its proper time\ \cite{K-M}. Moreover, by regarding both proper time and rest mass as operators, we were able to derive a limitation on the accuracy of a clock which coincides with that reached by another method\ \cite{SW}. Thus our approach leads to some desirable results, but at the same time we are faced with a fundamental difficulty: If we let an operator $T$ correspond to the coordinate time such that it satisfies the relation $[T, H]=i{\hbar }$ with the Hamiltonian $H$, then the spectrum of the Hamiltonian has to be continuous; this was proved by Pauli\ \cite{Pau}. This manifestly contradicts the existence of a discrete energy spectrum. Similarly, if we assume that the proper time $\tau $ and the rest mass $m$ are operators which satisfy the relation $[\tau, m]=i{\hbar }/c^2$, we are led to a result that conflicts with the existence of a discrete mass spectrum. 

Our objective in this letter is to propose an attempt to restate our problem in order to dispose of this contradiction. We focus our attention on a spin $1/2$ particle and try to find a quantum mechanical formulation of its proper time within the framework of the Dirac theory for relativistic quantum mechanics. We do not require any additional assumptions such that the rest mass be represented by 
an operator: It will be shown that an operator corresponds to the rate ${\dot \tau }\equiv d\tau /dt$ of the proper time and that the proper time itself can be represented through the expectation value of this operator. Further, it will be shown that ${\dot \tau }$ contains terms which oscillate with a very high frequency; this rapid oscillation is similar to the zitterbewegung which appears for the velocity of the particle. We can deduce, as an effect of these terms, the existence of an interference between the magnetic field and the rate of proper time. The magnitude of this interference is so small that we cannot detect its effect if we restrict our attention to experiments using present technology. However, several neutron stars called magnetars have been observed recently\ \cite{Ibra}; the magnetic field in the neighborhood of such a star is so strong that this interference becomes non-negligible. Therefore there is a possibility that the conclusion arrived at in this letter may find some application in future astrophysics. 

We consider a spin $1/2$ particle with finite mass $m$; the Dirac wave function is denoted by $\psi (x)$. We begin by finding a Hermitian matrix $D$ for which the integral 
\begin{equation}
\int _{t_0}^tdt\langle \psi \vert D\vert \psi \rangle 
=\int _{t_0}^tdt\int d^3{\bf x}\psi (x)^{\dagger }D\psi (x) \label{eq:Dnointegral}
\end{equation}
can be interpreted as \lq \lq the quantum mechanical proper time of the particle'' which passes in the course of an interval $[t_0, t]$ of coordinate time. The reason we take the integral (\ref{eq:Dnointegral}) as our starting point is as follows:
\begin{enumerate}
\item In the special theory of relativity, the proper time of a particle depends on the history of the particle, in the sense that it is determined by the orbit in space-time. On the other hand, the rate ${\dot \tau }$ does not depend on the history. 
\item In the quantum mechanics of a single particle, an operator can describe a physical quantity at $t={\rm constant}$ but cannot, however, represent any quantity which depends on the history of the particle. 
\item Therefore, if we could define \lq \lq the quantum mechanical proper time'' of a particle in the Dirac theory, it would be natural to think that the rate ${\dot \tau }$ can be represented by an operator, and that the proper time itself be given by the integral of the expectation value of the operator. 
\end{enumerate}
Now, in order to refer to \lq \lq the proper time'', the integral (\ref{eq:Dnointegral}) has to be invariant under all Lorentz transformations
\begin{equation}
x'^{\mu }={L^{\mu }}_{\nu }x^{\nu }. \label{eq:Lorentz}
\end{equation}
To require that this invariance be represented by the equation 
\begin{equation}
\int _{t_0}^tdt\int d^3{\bf x}\psi (x)^{\dagger }D\psi (x)=
\int _{t'_0}^{t'}dt'\int d^3{\bf x}'\psi '(x')^{\dagger }D\psi '(x')  \label{eq:demandori}
\end{equation}
is too formal. Equation (\ref{eq:demandori}) has no meaning because the interval $[t_0', t']$ cannot be uniquely determined from the interval $[t_0, t]$. Accordingly, the authors represent this condition by the equation 
\begin{equation}
\int _{\Omega }d^4x\psi (x)^{\dagger }D\psi (x)=\int _{\Omega '}d^4x'\psi '(x')^{\dagger }D\psi '(x'), \label{eq:okikaetajyoken}
\end{equation}
where $\Omega $ denotes an arbitrary domain in space-time and $\Omega '$ its image under the Lorentz transformation. The Jacobian of a Lorentz transformation is always equal to $1$, hence if we introduce an operator $\Lambda $ such that  
$$\psi '(x')=(\Lambda \psi )(x),$$
then Eq. (\ref{eq:okikaetajyoken}) is equivalent to  
\begin{equation}
\int _{\Omega }d^4x\psi (x)^{\dagger }D\psi (x)
=\int _{\Omega }d^4x\left( \Lambda \psi \right) (x)^{\dagger }D
(\Lambda \psi )(x). \label{eq:Dyousei} 
\end{equation}
In the following we show that we can determine the Hermitian matrix $D$ by using Eq. (\ref{eq:Dyousei}): Our notation for various matrices follows that of \ \cite{B-D} and we use units in which ${\hbar }=c=1$. For an infinitesimal Lorentz transformation
$$L_{\mu \nu }=\eta _{\mu \nu }+\omega _{\mu \nu }\hskip 1cm (\eta _{\mu \nu }\equiv {\rm diag.}(1, -1, -1, -1), \hskip 5mm \omega _{\mu \nu }=-\omega _{\nu \mu }),$$
we have (see p.21 in\ \cite{B-D}) 
$$(\Lambda \psi )(x)=\left( 1+{1\over 4}\gamma ^{\mu }\gamma ^{\nu }\omega _{\mu \nu }\right) \psi (x).$$
In this case the condition (\ref{eq:Dyousei}) implies that 
\begin{equation}
\omega _{\mu \nu }\left( D\gamma ^{\mu }\gamma ^{\nu }+
{\gamma ^{\nu }}^{\dagger }{\gamma ^{\mu }}^{\dagger }D\right) =0.
\label{eq:formugensyo}
\end{equation}
Using the expression 
$$\gamma ^0=\pmatrix{ I_2 & \cr & -I_2 \cr } \hskip 1cm \gamma ^j=\pmatrix{ & \sigma _j \cr -\sigma _j & \cr } \hskip 5mm (j=1, 2, 3)$$
for the matrices $\gamma ^{\mu }$, we can conclude from (\ref{eq:formugensyo}) that 
\begin{equation}
D=\pmatrix{ aI_2 & ibI_2 \cr -ibI_2 & -aI_2 \cr } ,\label{eq:abconstants}
\end{equation}
where $I_2$ denotes the unit matrix of degree $2$ and $a, b$ arbitrary real constants. When we take the time-reversal transformation
$$t'=-t, \hskip 5mm x'^j=x^j \hskip 15mm (j=1, 2, 3),$$
we have (see p.73 in\ \cite{B-D}) 
$$\left( \Lambda \psi \right) (x)=i\gamma ^1\gamma ^3\psi ^*(x).$$
In this case Eq. (\ref{eq:Dyousei}) implies that 
$$\gamma ^1\gamma ^3(^tD)\gamma ^1\gamma ^3=-D,$$
where $^tD$ denotes the transpose of the matrix $D$. This equation claims that the constant $b$ in Eq. (\ref{eq:abconstants}) has to be zero, that is, the matrix $D$ must be 
\begin{equation}
D=a\pmatrix{ I_2 & \cr & -I_2 \cr }=a\beta \hskip 1cm (\beta \equiv \gamma ^0). \label{eq:Dnokatachi}
\end{equation}
If we take the space reflection 
$$t'=t, \hskip 5mm x'^j=-x^j \hskip 15mm (j=1, 2, 3)$$
as the Lorentz transformation (\ref{eq:Lorentz}), we have (see p.25 in\ \cite{B-D}) 
$$\left( \Lambda \psi \right) (x)=e^{i\theta }\gamma ^0\psi (x).$$
In this case Eq. (\ref{eq:Dyousei}) is equivalent to the equation $\gamma ^0D\gamma ^0=D$, which is automatically satisfied if $D$ has the form indicated by (\ref{eq:Dnokatachi}). 
Thus we have shown that $\tau (t)$ defined by 
\begin{equation}
\tau (t)\equiv \int _{t_0}^tdt\langle \psi \vert \beta \vert \psi \rangle 
=\int _{t_0}^tdt\int d^3{\bf x}\psi (x)^{\dagger }\beta \psi (x)
\label{eq:taunoteigi}
\end{equation}
is invariant under any Lorentz transformation; the precise meaning is that Eq. (\ref{eq:okikaetajyoken}) is satisfied for any domain $\Omega $ and for all Lorentz transformations including time-reversal and space reflection. 

Now, our next step is to show the basis on which the authors judge that $\tau (t)$ given by (\ref{eq:taunoteigi}) really represents the proper time of a particle in the Dirac theory. In order to do that, we focus on the rate 
\begin{equation}
{\dot \tau }(t)={{d\tau }\over {dt}}=\langle \psi \vert \beta \vert \psi \rangle .  \label{eq:okureope}
\end{equation}
For a free particle, we have 
$$\psi (t, {\bf x})=e^{-iHt}\psi _0({\bf x}) \hskip 1cm (\psi _0({\bf x})\equiv \psi (0, {\bf x}))$$
and 
\begin{equation}
{\dot \tau }(t)=\langle \psi _0\vert \beta (t)\vert \psi _0\rangle ,  \label{eq:expectation}
\end{equation}
where $H\equiv {\bf \alpha }\cdot {\bf p}+m\beta $ denotes the Dirac Hamiltonian and we set $\beta (t)=e^{iHt}\beta e^{-iHt}.$ We can easily show that  
$$e^{-iHt}=e^{-i\vert H\vert t}+i\left( 1-H/\vert H\vert \right) \sin \vert H\vert t \hskip 15mm \left( \vert H\vert \equiv {\sqrt {{\bf p}^2+m^2}} \right) ,$$
and therefore we have
\begin{eqnarray}
\beta (t)&=&\beta -2iP\sin (\vert H\vert t)\beta e^{-i\vert H\vert t}+2ie^{i\vert H\vert t}\beta \sin (\vert H\vert t)P \nonumber \\
&&\hskip 3cm +4P\sin (\vert H\vert t)\beta \sin (\vert H\vert t)P, \label{eq:Zitter}
\end{eqnarray}
where $P\equiv (1/2)\left( 1-H/\vert H\vert \right) $ is the projection operator onto the space of states with negative energy. Each term except $\beta $ in the right-hand side of (\ref{eq:Zitter}) oscillates with a very high frequency $\vert H\vert /{\hbar }>mc^2/{\hbar }$\  ($\approx 10^{21} {\rm Hz}$ for an electron). Moreover, if $\psi _0$ is a superposition of states with positive energy, then the expectation value (\ref{eq:expectation}) coincides with the value of 
$\langle \psi _0\vert \beta \vert \psi _0\rangle $ because of the existence of the projection operator $P$. This value, furthermore, is equal to 
$$\langle \psi _0\vert m/{\vert H\vert }\vert \psi _0\rangle $$
because we have the equation
$${{1+H/{\vert H\vert }}\over 2}\beta {{1+H/{\vert H\vert }}\over 2}={m\over {\vert H\vert }}{{1+H/{\vert H\vert }}\over 2},$$
where $(1+H/\vert H\vert )/2$ is the projection operator onto the space of states with positive energy. When the value of $\vert H\vert $ equals $m/{\sqrt {1-v^2}},$ we have the familiar factor $m/{\vert H\vert }={\sqrt {1-v^2}}$ where $v$ denotes the velocity of the particle. That is to say, for a free particle with a definite positive energy and with a definite momentum, we have 
$${\dot \tau }(t)={\sqrt {1-v^2}}\times \langle \psi _0\vert \psi _0\rangle .$$
In this sense, $\tau (t)$ defined by (\ref{eq:taunoteigi}) extends 
the classical concept of proper time to the Dirac theory for relativistic quantum mechanics. Judging from the above result and the Lorentz invariance (\ref{eq:okikaetajyoken}), the authors think that we have found a quantum theoretical representation for the proper time of spin $1/2$ particles. The formula (\ref{eq:okureope}) means that the operator $\beta $ corresponds to the rate of the proper time. 

Our next step is to show a new phenomenon which can be predicted through Eq. (\ref{eq:okureope}). Let us consider the case of an electron in a prescribed external electromagnetic field. The Hamiltonian is 
$$H={\bf \alpha }\cdot ({\bf p}-e{\bf A})+e\phi +m\beta .$$
In the following, we use a procedure developed by Foldy and Wouthuysen\ \cite{FW} which decouples the Dirac equation into two two-component equations; one reduces to the Pauli description in the nonrelativistic limit and the other describes the negative-energy states. The unitary transformation 
$$U\equiv e^{\beta {\bf \alpha }\cdot ({\bf p}-e{\bf A})/2m}$$
transforms (see p.51 in\ \cite{B-D}) the Hamiltonian $H$ into 
$$H'\equiv UHU^{\dagger }={1\over {2m}}\beta ({\bf p}-e{\bf A})^2+e\phi +m\beta -{e\over {2m}}\beta {\bf \sigma }\cdot {\bf B},$$
where we keep terms up to order $v^2$ assuming that ${\bf \alpha }\cdot ({\bf p}-e{\bf A})/m=O(v)$ and $e\phi /m=O(v^2)$ in the nonrelativistic limit. The upper and the lower half of $\psi '(x)\equiv U\psi (x)$ are denoted by $\Phi '(x)$ and $\chi '(x)$ respectively; 
$$\psi '(x)\equiv U\psi (x)=\pmatrix{ \Phi '(x) \cr \chi '(x) \cr }.$$
Using the fact that the operator $H'$ does not couple the components $\Phi '(x)$ and $\chi '(x)$, we can show that $\Phi '$ is large compared with $\chi '$ if $\psi $ is a superposition of only positive-energy states; we then have 
\begin{equation}
\chi '/\Phi '=O(v^3). \label{eq:ls}
\end{equation}
We get the Pauli Hamiltonian
$${1\over {2m}}({\bf p}-e{\bf A})^2+e\phi -{e\over {2m}}{\bf \sigma }\cdot {\bf B}$$
when we restrict the action of $H'$ to the large component $\Phi '$ and eliminate $m$ which corresponds to the rest energy. Now, let us examine the right side of (\ref{eq:okureope}) for $\psi $ which is a superposition of positive-energy states. We have, to order $v^2$, 
\begin{equation}
U\beta U^{\dagger } =\beta -{1\over m}{\bf \alpha }\cdot ({\bf p}-e{\bf A}) -{1\over {2m^2}}\beta ({\bf p}-e{\bf A})^2+{e\over {2m^2}}\beta {\bf \sigma }\cdot {\bf B} \label{eq:betadash} .
\end{equation}
The second term ${\bf \alpha }\cdot ({\bf p}-e{\bf A})/m$ in (\ref{eq:betadash}) couples large and small components, and we have ${\bf \alpha }\cdot ({\bf p}-e{\bf A})/m=O(v)$ as mentioned above, therefore we get  
$$\langle \psi '\vert {\bf \alpha }\cdot ({\bf p}-e{\bf A})/m\vert \psi '\rangle =O(v^4)$$
from Eq. (\ref{eq:ls}); we can neglect this term. Other terms in (\ref{eq:betadash}) do not couple $\Phi '$ and $\chi '$, hence we have \begin{equation}
{\dot \tau }(t)=\langle \psi (x)\vert \beta \vert \psi (x)\rangle =\langle \Phi '\vert 1-{1\over {2m^2}}({\bf p}-e{\bf A})^2+{e\over {2m^2}}{\bf \sigma }\cdot {\bf B} \vert \Phi '\rangle  \label{eq:betatrans}
\end{equation}
to order $v^2$ (see\ \cite{order}). The operator $1-({\bf p}-e{\bf A})^2/{2m^2}$ in (\ref{eq:betatrans}) corresponds to a non-relativistic approximation of the factor ${\sqrt {1-v^2}}$. The third term 
\begin{equation}
{e\over {2m^2}}{\bf \sigma }\cdot {\bf B} \label{eq:thirdterm}
\end{equation}
is very important for us. It means that the interaction between the spin of the particle and the external magnetic field has an influence on the speed of evolution of the proper time. That is, the evolution of the proper time is affected not only by the velocity of the particle but also by its spin state. This new effect, of course, has no correspondence in the special theory of relativity.

Finally, we have to estimate the magnitude of the value of (\ref{eq:thirdterm}) and discuss the physical meaning of our result. Assuming that the magnetic flux density ${\bf B}$ is measured in units of teslas, we can show that 
$$\vert  {e\over {2m^2}}{\bf \sigma }\cdot {\bf B}\vert \approx 2\times 10^{-10}\vert {\bf B}\vert .$$
Therefore, as far as laboratory experiments are concerned, the new effect is extremely small. However, a few neutron stars with a very intense magnetic field have been recently observed\ \cite{Ibra}; such stars are called magnetars, and the strength of their magnetic field is of order $10^{10} {\rm T}$ or more. This means that the effect of the term (\ref{eq:thirdterm}) becomes important for electrons around such a star. Hence there is a possibility that our new term has interesting effects in some astrophysical phenomena. It is to be hoped that this subject will be investigated further.  

In this letter we have found a quantum theoretical formulation of proper time for spin $1/2$ particles within the framework of the Dirac theory. It has been shown that the evolution of the proper time is affected not only by the velocity of the particle but also by its spin state. This interference between the magnetic field and the rate of proper time may have its effects in astrophysical phenomena.   

The authors would like to thank Dr. Yuko Motizuki for useful comments.

\vfill

\begin{thebibliography}{99}
\bibitem{order} The restriction of the action of $\partial H'/\partial m$ to the large component $\Phi '$ coincides with the operator on the right side of (\ref{eq:betatrans}). This is true even when we consider $H'=UHU^{\dagger }$ and ${\dot \tau }$ keeping terms up to order $v^4$. On the other hand, we know that we are led to the equation ${\dot \tau }=\partial [{\rm Hamiltonian}]/\partial m$ if we assume that $\tau $ and $m$ are operators which satisfy the relation $[\tau, m]=i$\ \cite{Gr, K-M}. Various aspects of this analogy remain insufficiently explored.
\end{thebibliography}
\end{document}